\documentclass[aps,pra,preprint,superscriptaddress,longbibliography]{revtex4-1}

\usepackage{amsmath}
\usepackage{amsthm}
\usepackage{amsfonts}
\usepackage{amssymb}
\usepackage{xcolor,graphicx}
\usepackage{tikz}
\usepackage{color}
\usepackage[pdftex]{hyperref} 
\usepackage{longtable}
\usepackage{array}
\usepackage{dsfont}

\begin{document}
	
\title{Underlying SUSY in a generalized Jaynes-Cummings model}

\author{F. H. Maldonado-Villamizar}
\email[e-mail: ]{fmaldonado@inaoep.mx}
\affiliation{CONACYT-Instituto Nacional de Astrof\'isica, \'Optica y Electr\'onica, Calle Luis Enrique Erro No. 1, Sta. Ma. Tonantzintla, Pue. CP 72840, M\'exico} 

\author{C. A. Gonz\'alez-Guti\'errez}
\affiliation{ Instituto de Investigaciones en Matem\'aticas Aplicadas y en Sistemas, Universidad Nacional Aut\'onoma de M\'exico, Ciudad Universitaria, Ciudad de M\'exico 04510, M\'exico}

\author{L. Villanueva-Vergara}
\affiliation{Instituto Nacional de Astrof\'isica, \'Optica y Electr\'onica, Calle Luis Enrique Erro No. 1, Sta. Ma. Tonantzintla, Pue. CP 72840, M\'exico}

\author{B. M. Rodr\'iguez-Lara}
\email[e-mail: ]{bmlara@tec.mx}
\affiliation{Tecnologico de Monterrey, Escuela de Ingenier\'ia y Ciencias, Ave. Eugenio Garza Sada 2501, Monterrey, N.L., 64849, M\'exico}

\begin{abstract}
	We propose a generalized Jaynes-Cummings model that includes but is not limited to an extensive collection of experimental and theoretical proposals from the literature. 
	It covers nonlinear boson terms, nonlinear dispersive and multi-boson exchange interaction. 
	Our model features an underlying Lie graded algebra symmetry reminiscent to supersymmetric quantum mechanics. 
	This allows us to propose a diagonalization scheme and calculate its analytic time evolution.  
	In consequence, we are able to construct closed forms for relevant observables and explore the dynamics of particular realizations of our model independent of their complexity.
	As an practical example, we show the evolution of the population inversion and the boson quadratures for an initial state consisting of the qubit in the ground state interacting with a coherent field for a selection of cases including the standard JC model with Stark shift, Kerr-like terms, intensity dependent coupling, multi-boson exchange and algebraic deformations. 
\end{abstract}

	
	\maketitle
	
\section{Introduction}

The distinction between bosons and fermions is closely related to the statistics they follow \cite{Landau_1980p158}.
Each behave in an unique manner but both follow an underlying supersymmetry (SUSY) \cite{Neveu_1971p86,Ramond_1971p2415} relating a set of bosonic $\left[\hat{a},\hat{a}^{\dagger}\right]=1$ and fermionic $\left\{\hat{f}_{-},\hat{f}_{+}\right\} = 1$ degrees of freedom.
In its simplest form \cite{Witten_1981p513}, SUSY  requires a set of constrains,  
\begin{equation}\label{eq:001}
\hat{H}=\left\{\hat{Q},\hat{Q}^{\dagger}\right\}, ~ \left[\hat{H},\hat{Q}\right]=\left[\hat{H},\hat{Q}^{\dagger}\right]=0, ~ \hat{Q}^{\dagger 2}=\hat{Q}^{2}=0,
\end{equation}
relating the so-called SUSY-Hamiltonian $\hat{H}$ with the SUSY-charges $\hat{Q}$ and $\hat{Q}^{\dagger}$.
In general, it is possible to extend this construction in a manner where each added bosonic degree of freedom induces a fermionic one \cite{Decrombrugghe_1983p99}.

The simplest example of SUSY implies a bosonic and a fermionic degrees of freedom \cite{Saolomonson_1982p509}
For the sake of experimental feasibility, we think of a qubit interacting with a bosonic degree of freedom; for example, the two-internal levels of a neutral atom interacting with a single mode of the quantum electromagnetic field \cite{Tanoudji_1998p707}, those of a trapped ion interacting with a quantum center of mass vibration mode \cite{Blatt_2008p1008}, or a superconducting qubit interacting with the quantum mode of a strip-line resonator \cite{Niemczyk2010p772}.
In this case, the SUSY Hamiltonian and charges have the form,
\begin{eqnarray}
\hat{H} = \hat{\sigma}_{+} \hat{\sigma}_{-} \hat{a}^{\dagger} \hat{a} +  \hat{\sigma}_{-} \hat{\sigma}_{+} \hat{a} \hat{a}^{\dagger}, ~
\hat{Q}= \hat{\sigma}_{+} \hat{a}^{\dagger}, ~ 
\hat{Q}^{\dagger} = \hat{\sigma}_{-} \hat{a},
\end{eqnarray}
in terms of up (down) Pauli operators for the pseudo-fermion $\hat{\sigma}_{+}$ ($\hat{\sigma}_{-}$) and creation (annihilation) operators for the boson mode $\hat{a}^{\dagger}$ ($\hat{a}$).
It is straightforward to see that the spectrum of this SUSY Hamiltonian is given by $E_{e,n} = n$ with $\vert \psi \rangle = \vert e, n \rangle$ and $E_{g,n} = n+1$ with $\vert \psi \rangle = \vert g, n \rangle$, where the notation $\vert g \rangle$ ($\vert e \rangle$) refers to the ground (excited) state of the qubit and $\vert n \rangle$ to a Fock state of the boson field. 
This type of SUSY Hamiltonians are convoluted but possible to realize in experiments; for example, squeezed states relating an interaction free qubit and boson field with the anti-Jaynes-Cummings model in the strong coupling regime \cite{Orszag_1988pL1059}.

Here, we want to focus on the fact that this SUSY formulation allows for two  auxiliary charges
\begin{eqnarray}
\hat{Q}_{X} = \hat{Q}^{\dagger} + \hat{Q} \quad \mathrm{and} \quad \hat{Q}_{Y} = -i \left( \hat{Q}^{\dagger} - \hat{Q} \right),
\end{eqnarray}
that are the square root of the SUSY Hamiltonian,
\begin{eqnarray}
\hat{Q}_{X}^{2} = \hat{Q}_{Y}^{2} = \hat{H},
\end{eqnarray}
these auxiliary SUSY charges are straightforward to realize as the Jaynes-Cummings model \cite{Jaynes_1963p89} in experimental setups involving the resonant interaction between a the internal levels of a neutral Rydberg atom and a quantized electromagnetic field \cite{Haroche_1985p347}, two internal levels of a trapped ion interacting with the quantized motion of its centre of mass \cite{Blatt_2008p1008}, or the interaction of a superconducting qubit with a strip-line resonator \cite{Wallraff_2004p162}.

In the following, we explore a generalization of the Jaynes-Cummings model that includes non-linear, multi-boson exchange, Sec. \ref{sec:Sec2}, and its relation to an algebra with SUSY characteristics, Sec. \ref{sec:Sec3}.
Then, in Sec. \ref{sec:Sec4}, we diagonalize the system using the structure of our proposed algebra and present the time evolution for observables of interest.
These analytic results allows us to visualize the dynamics of specific realizations of our model that have been discussed through the years in Sec. \ref{sec:Sec5}.
Finally, we close with our conclusion in Sec. \ref{sec:Sec6}.

\section{Generalized Jaynes-Cummings model} \label{sec:Sec2}
The introduction of the Jaynes-Cummings model (JC) \cite{Jaynes_1963p89} to describe the interaction of a two-level system with a boson field under the rotating wave approximation (RWA) opened the door to more complicated models from both the theoretical and experimental perspectives. 
We focus on a generalized JC model, 
\begin{equation}
\hat{H} =  \omega \hat{n} +\frac{1}{2} \omega_{0} \hat{\sigma}_{z}+  \hat{\sigma}_{z} F(\hat{n}) + G(\hat{n}) + g \left[ \hat{\sigma}_{+} \hat{a}^{k}f(\hat{n})  + \hat{\sigma}_{-} f(\hat{n})\hat{a}^{\dagger k} \right],
\end{equation}
that accounts for families of reported models and more.
Here, the frequency $\omega_{0}$ provides the qubit energy gap and the up (down) and population inversion operators, $\hat{\sigma}_{+}$ ($\hat{\sigma}_{-}$) and $\hat{\sigma}_{z}$, provide its dynamics. 
The boson field frequency is $\omega$ with creation (anihilation) and boson excitation number operators, $\hat{a}^{\dagger}$ ($\hat{a}$) and $\hat{n}$, in that order.
The first two terms in the right hand side are the free energy of the qubit and the boson field. 
The third term implies nonlinear shifting of the spectrum as a function of the boson excitation number; it includes the Stark effect.
The fourth term is a collection of nonlinear effects in the boson field; it includes the Kerr effect.
The fifth term is the nonlinear, multi-boson interaction between the qubit and the boson field under the rotating wave approximation.

Our model covers but is not limited to a cohort of examples from the literature. 
The obvious one is the  Jaynes-Cummings (JC) model  \cite{Jaynes_1963p89} describing the interaction of a qubit with a boson field in the RWA and, as discussed in the introduction, relates to standard supersymmetric quantum mechanics. 
One of the first extensions of the JC model used an intensity dependent coupling and multiboson exchange interaction \cite{Singh_1982p3206,Sukumar_1984p885}. 
Soon after, a slight modification included nonlinear effects such as Kerr-like terms and two-boson exchange \cite{Alsing_1987p177,Nasreen_1993p1292,Obada_1998p289,Werner_1991p4623}. 
In  these works, photon statistics and time evolution of physical observables were presented. 
The addition of the Stark shift, an interesting effect describing the qubit energy gap dependence on the intensity of the field, came later  \cite{Puri_1988p2021,Joshi_1992p5056}.  
Nonlinear extensions for the occupation number were proposed as a generalization to the Kerr effect \cite{Buzek_1990p4079}.
Then, trapped ions were proposed to realize nonlinear multiboson exchange interaction \cite{Vogel_1995p4214}.
An algebraic generalizaton was proposed to study coherent states for an anharmonic perturbation to the JC model. 
Some of us studied a, slightly complicated in hindsight, generalization \cite{Rodriguez-Lara_2005p023811,Rodriguez-Lara_2013p12888,Rodriguez-Lara_2014p1719,Rodriguez-Lara_2013p87,Ramos-Prieto_p1560005,Soto-Eguibar_2014p1335} that reduces to our general scheme in the following section.

\section{Graded Lie algebra} \label{sec:Sec3}

Let us focus on just the interaction part of our generalized Jaynes-Cummings Hamiltonian and recast it into the form,
\begin{align}
\hat{H}_{I} = g \left( \hat{\mathcal{Q}}^{\dagger} + \hat{\mathcal{Q}} \right),
\end{align}
where we define the nilpotent charges, 
\begin{align}
\hat{\mathcal{Q}}^{\dagger} = \hat{\sigma}_{+} f(\hat{n}) \hat{a}^{k}, \quad \mathrm{and} \quad \hat{\mathcal{Q}} = \hat{\sigma}_{-}  \hat{a}^{\dagger k} f(\hat{n}),
\end{align}
such that $\hat{\mathcal{Q}}^{\dagger 2} = \mathcal{Q}^{2} = 0 $.
These provide the SUSY Hamiltonian,
\begin{align}
\hat{\mathcal{H}} = & \left\{ \hat{\mathcal{Q}}^{\dagger} , \hat{\mathcal{Q}} \right\}, \\
=&  \hat{\sigma}_{+}\hat{\sigma}_{-} \hat{a}^{k}\hat{a}^{\dagger k}f^{2}(\hat{n}) + \hat{\sigma}_{-}\hat{\sigma}_{+} \hat{a}^{\dagger k}\hat{a}^{k}f^{2}(\hat{n}-k),
\end{align}
that commutes with the charges $ \left[ \hat{\mathcal{H}}, \hat{\mathcal{Q}}^{\dagger} \right] = \left[ \hat{\mathcal{H}}, \hat{\mathcal{Q}} \right] = 0$, and whose diagonal elements are isospectral sectors,
\begin{align}\label{eq:0015}
\begin{split}
\hat{\mathcal{H}}_{F} \vert e, n \rangle = & \hat{\mathcal{Q}}^{\dagger}  \hat{\mathcal{Q}} \vert e, n \rangle = n f^{2}(n-1) \vert e, n \rangle, \\
\hat{\mathcal{H}}_{B} \vert g, n \rangle = & \hat{\mathcal{Q}} \hat{\mathcal{Q}}^{\dagger} \vert g, n \rangle = (n+1) f^{2}(n) \vert g, n \rangle,
\end{split}
\end{align}
connected by the intertwining relations,
\begin{align}
\hat{\mathcal{Q}} \hat{\mathcal{H}}_{F} = \hat{\mathcal{H}}_{B} \hat{\mathcal{Q}}, \qquad \mathrm{and} \qquad
\hat{\mathcal{H}}_{F} \hat{\mathcal{Q}}^{\dagger} = \hat{\mathcal{Q}}^{\dagger} \hat{\mathcal{H}}_{B}.
\end{align}
It is possible to define two auxiliary charge operators, 
\begin{align}
\hat{\mathcal{Q}}_{X} = \hat{\mathcal{Q}}^{\dagger} + \hat{\mathcal{Q}} \qquad \mathrm{and} \qquad
\hat{\mathcal{Q}}_{Y} = i \left( \hat{\mathcal{Q}}^{\dagger} - \hat{\mathcal{Q}} \right),
\end{align}
that are the square root of the Hamiltonian $\hat{\mathcal{Q}}_{X}^{2} = \hat{\mathcal{Q}}_{Y}^{2} = \mathcal{H}$.
Thus, the interaction part of our generalized Jaynes-Cummings model is proportional to the square root of the Hamiltonian $\hat{\mathcal{H}}$ with an underlying SUSY algebra \cite{Andreev_1989p507,Fan_1995p221,LUHuai-Xin_2000p319}.

Now, let us recast our complete Hamiltonian, 
\begin{equation}\label{eq:014}
\hat{H} = \omega \left( \hat{\mathcal{N}} - \hat{\mathcal{B}} \right) + \frac{\omega_{0}}{k} \hat{\mathcal{B}} +  F(\hat{\mathcal{N}} - \hat{ \mathcal{B}}) \hat{ \mathcal{B}} + G \left( \hat{\mathcal{N}} - \hat{\mathcal{B}} \right) + g  \left( \hat{\mathcal{Q}}^{\dagger} + \hat{\mathcal{Q}} \right),
\end{equation}
where we define the total excitation number, 
\begin{align}
\hat{\mathcal{N}} = \hat{n} + \hat{\mathcal{B}},
\end{align}
in terms of the boson number operator $\hat{n}$ and the scaled Pauli z-matrix,
\begin{align} 
\hat{\mathcal{B}}= k \hat{\sigma}_{z} / 2 .
\end{align}
We assume that the nonlinear boson functions are continuous and differentiable, such that $F(x)= \sum_{j} F_{j}x^{j}$ and $G(x)= \sum_{j} G_{j}x^{j}$ with 
the shorthand notation $f_{j}(0) = d^{j} f(x) / dx^{j}  \vert_{x=0}$.
Both the SUSY Hamiltonian and the total excitation number commute with all other operators involved in our model, 
\begin{eqnarray}
\left[ \hat{\mathcal{O}}_{j} , \hat{\mathcal{H}} \right] = \left[ \hat{\mathcal{O}}_{j} , \hat{\mathcal{N}} \right]   = 0,
\end{eqnarray}
where the place holder operator $\hat{O}_{j}$ stands for elements of the set $ \hat{\mathcal{O}} = \left\{ \hat{\mathcal{Q}},  \hat{\mathcal{Q}}^{\dagger}, \hat{\mathcal{H}}, \hat{\mathcal{N}}, \hat{\mathcal{B}}\right\}$.
The commutation relations between the charges and the scaled Pauli-z operator,
\begin{align}
\left[ \hat{\mathcal{B}}, \hat{\mathcal{Q}}^{\dagger} \right] =  k \hat{\mathcal{Q}}^{\dagger}, \quad
\left[ \hat{\mathcal{B}}, \hat{\mathcal{Q}} \right] = -  k \hat{\mathcal{Q}}, \quad
 \left[ \hat{\mathcal{Q}}^{\dagger}, \hat{\mathcal{Q}}\right] = \frac{2}{k} \hat{\mathcal{H}}  \hat{\mathcal{B}},
\end{align}
are reminiscent of a deformed $su(2)$ algebra.
These relations will come handy in the diagonalization of our model.

\section{Diagonalization} \label{sec:Sec4}

It is possible to use the properties of the scaled Pauli-z operator to recast our generalized Jaynes-Cummings Hamiltonian in the form, 
\begin{align}
\hat{H} = \omega \hat{\mathcal{N}} + \mathcal{F} (\hat{\mathcal{N}}) + \left[ \frac{\omega_{0}}{k} - \omega + \mathcal{G}(\hat{\mathcal{N}})  \right] \hat{\mathcal{B}} + g \left( \hat{\mathcal{Q}}^{\dagger} + \hat{\mathcal{Q}}^{\dagger} \right), \label{eq:SUSYRecast}
\end{align}
where the auxiliary functions in terms fo the total excitation number relate to the nonlinear boson number functions in the following manner,
\begin{widetext}
\begin{align}
\mathcal{F} (\hat{\mathcal{N}}) =& \sum_{j=0}^{\infty}\sum_{s=0}^{j}\binom{j}{2s}\left(\dfrac{k}{2}\right)^{2s}G_{j}\mathcal{\hat{N}}^{j-2s}  -\sum_{j=0}^{\infty}\sum_{s=0}^{j}\binom{j}{2s+1}\left(\frac{k}{2}\right)^{2s-1}F_{j}\mathcal{\hat{N}}^{j-2s-1}  ,\\
\mathcal{G}(\hat{\mathcal{N}}) =&-\sum_{j=0}^{\infty}\sum_{s=0}^{j}\binom{j}{2s+1}\left(\frac{k}{2}\right)^{2s}G_{j}\mathcal{\hat{N}}^{j-2s-1} + \sum_{j=0}^{\infty}\sum_{s=0}^{j} \binom{j}{2s}\left(\dfrac{k}{2}\right)^{2s-1}F_{j}\mathcal{\hat{N}}^{j-2s}.
\end{align}
\end{widetext}
Here, we used the fact that $ \hat{\mathcal{B}}^{2 j} = ( k /2)^{2j}$ and $\hat{\mathcal{B}}^{2 j + 1} = (k/2)^{2j} \hat{\mathcal{B}}$.
As the total number excitation is an invariant of the model, the first two terms in the right hand side of Eq.(\ref{eq:SUSYRecast}) only introduce a phase factor. 
We move into a rotating frame defined by these terms,
\begin{align} 
	\vert \psi \rangle = e^{-i \left[ \omega \hat{\mathcal{N}} + \mathcal{F} (\hat{\mathcal{N}})  \right]t} \vert \phi \rangle,
\end{align} 
such that we obtain an effective Hamiltonian,
\begin{align}
\hat{H}_{\phi} = \left[ \frac{\omega_{0}}{k} - \omega + \mathcal{G}(\hat{\mathcal{N}}) \right] \hat{\mathcal{B}} + g \left( \hat{\mathcal{Q}}^{\dagger} + \hat{\mathcal{Q}}^{\dagger} \right),
\end{align}
where the factor accompanying the scaled Pauli-z operator $\hat{\mathcal{B}}$ commutes with all other operators. 

Now, we move into the frame introduced by the displacement transformation, 
\begin{align}
\hat{D}(\hat{\beta}) = e^{ - \frac{\hat{\beta}}{2} \hat{\mathcal{H}}^{-\frac{1}{2}} \left( \hat{\mathcal{Q}}^{\dagger} - \hat{\mathcal{Q}} \right) },
\end{align}
where the operator $\hat{\mathcal{H}}^{1/2}$ is the square root of elements in Eq.(\ref{eq:0015}). 
The effective Hamiltonian is diagonal in this frame,
\begin{align}\nonumber
\hat{H}_{D} =& \hat{D}^{\dagger}(\hat{\beta}) \hat{H} \hat{D}(\hat{\beta}), \\ = &  \left\{ \left[ \frac{\omega_{0}}{k} - \omega + \mathcal{G}(\hat{\mathcal{N}}) \right] \cos \hat{\beta}   +  \frac{2g}{k}\hat{\mathcal{H}}^{1/2} \sin \hat{\beta} \right\}  \hat{\mathcal{B}},
\end{align}
for a displacement parameter operator fulfilling,
\begin{align}
\tan  \hat{\beta}  = \frac{2g}{k}\hat{\mathcal{H}}^{1/2} \left[ \frac{\omega_{0}}{k} - \omega + \mathcal{G}(\hat{\mathcal{N}}) \right]^{-1}.
\end{align}
All involved terms are diagonal in the qubit and Fock basis and we used the expressions, 
\begin{align}
\hat{D}^{\dagger}(\hat{\beta}) \hat{\mathcal{B}} \hat{D}(\hat{\beta}) =&~ \hat{\mathcal{B}} \cos \hat{\beta}  - \dfrac{k}{2\hat{\mathcal{H}}^{1/2}}\left( \hat{\mathcal{Q}}^{\dagger} + \hat{\mathcal{Q}} \right) \sin \hat{\beta} , \nonumber\\
\hat{D}^{\dagger}(\hat{\beta}) \left( \hat{\mathcal{Q}}^{\dagger} + \hat{\mathcal{Q}} \right) \hat{D}(\hat{\beta}) =&~  \left( \hat{\mathcal{Q}}^{\dagger} + \hat{\mathcal{Q}} \right) \cos \hat{\beta}  + 2\hat{\mathcal{H}}^{1/2} \hat{\mathcal{B}} \sin  \hat{\beta} . \nonumber
\end{align}

\section{Eigenstates and time evolution}
In the original frame, it is possible to calculate the eigenstates of our model in terms of the manifold $\left\{ \vert{e, n} \rangle, \vert{g, n + k} \rangle \right\}$ with total excitation number $\mathcal{N} = \langle \hat{N} \rangle = n + k/2$,
\begin{align}\nonumber
\vert + , \mathcal{N} \rangle = &~ \hat{D}(\hat{\beta}) \vert e, n \rangle\\\nonumber  =&~ \cos\left(\dfrac{\beta(\mathcal{N})}{2}\right) \vert e,n \rangle +  \sin \left(\dfrac{\beta(\mathcal{N})}{2}\right)  \vert g, n+k\rangle , \\\nonumber
\vert - , \mathcal{N} \rangle = &~ \hat{D}(\hat{\beta}) \vert g, n + k \rangle\\ = &~ -  \sin\left(\dfrac{\beta(\mathcal{N})}{2}\right)   \vert e,n \rangle + \cos\left( \dfrac{\beta(\mathcal{N})}{2}\right)  \vert g, n+k\rangle , 
\end{align}
up to a common phase $\phi(\mathcal{N}) = \omega \left( n + k/2 \right) + F( n + k/2)$ and the relation
\begin{align}
\tan \beta(\mathcal{N}) = &~  \frac{2g}{k}\sqrt{\frac{\left(\mathcal{N}+\frac{k}{2}\right)!}{\left(\mathcal{N}-\frac{k}{2}\right)!}}f\left(\mathcal{N}-\frac{k}{2}\right) \left[ \frac{\omega_{0}}{k} - \omega + \mathcal{G}\left(\mathcal{N}\right) \right]^{-1}.
\end{align}
The corresponding eigenvalues,
\begin{align} 
	E_{\pm}(\mathcal{N},k) =\pm\Omega\left(\mathcal{N}\right),
\end{align}
involve a generalized Rabi frequency, 
\begin{equation}\label{eq:0028}
\Omega^{2}(\mathcal{N})=\left[\dfrac{\omega_{0}}{k}-\omega+\mathcal{G}\left(\mathcal{N}\right)\right]^{2}+\frac{4g^{2}}{k^{2}}\frac{\left(\mathcal{N}+\frac{k}{2}\right)!}{\left(\mathcal{N}-\frac{k}{2}\right)!}~f^{2}\left(\mathcal{N}-\frac{k}{2}\right).
\end{equation}

These results yield a time evolution in the diagonal frame, 
\begin{align}
\hat{U}(t) = e^{-i \left\{ \left[ \frac{\omega_{0}}{k} - \omega + \mathcal{G}(\hat{\mathcal{N}}) \right] \cos  \hat{\beta}  +  \frac{2g}{k}\hat{\mathcal{H}}^{1/2} \sin  \hat{\beta} \right\}  \hat{\mathcal{B}} t},
\end{align}
that helps us calculate the evolution of the Pauli-z operator
\begin{widetext}
\begin{align}\nonumber
\langle \hat{\sigma}_{z}(t)  \rangle&=\langle \psi (0)\vert D(\hat{\beta})\left[\hat{\sigma}_{z} \cos \hat{\beta}  + \dfrac{1}{\hat{\mathcal{H}}^{1/2}}\left( \hat{\mathcal{Q}}^{\dagger}e^{ik\hat{H}_{D}\hat{\mathcal{B}}^{-1}t} + \hat{\mathcal{Q}}e^{-ik\hat{H}_{D}\hat{\mathcal{B}}^{-1}t} \right) \sin \hat{\beta}\right]D^{\dagger}(\hat{\beta}) \vert \psi(0) \rangle.\\
\end{align}
\end{widetext}
For example, assuming an initial state with the qubit in the excited state and the boson field in a Fock state,
\begin{align}
\vert \psi(0) \rangle = \vert g, n\rangle,
\end{align}
it is straightforward to calculate,
\begin{align}\label{eq:IP}
\langle \hat{\sigma}_{z}\rangle_{\mathcal{N}}&= \cos^{2}\beta(\mathcal{N})+\sin^{2}\beta(\mathcal{N})\cos\left[\Omega(\mathcal{N})t\right].
\end{align}
The other observable, the boson field excitation number, is trivial,
\begin{eqnarray}
\langle \hat{n} \rangle = \langle \hat{\mathcal{N}} \rangle - \frac{k}{2} \langle \hat{\sigma}_{z} \rangle.
\end{eqnarray}
We use these expressions to compare several models included in our generalized Hamiltonian involving an initial state with the qubit in the ground state and the boson in a coherent state,
\begin{equation}
	\vert \psi(0)\rangle=\vert g, \alpha\rangle=\sum_{j=0}^{\infty}\frac{e^{-\frac{\vert \alpha\vert^{2}}{2}}}{\sqrt{j!}}\alpha^{j}\vert g, j\rangle.
\end{equation}
The evolution of the Pauli-z operator is,
\begin{equation}\label{eq:EvCoh}
\langle g,\alpha\vert\hat{\sigma}_{z}\vert g, \alpha\rangle=\sum_{j=0}^{\infty}\frac{e^{-\vert \alpha\vert^{2}}}{j!}\vert\alpha\vert^{2j}\langle \hat{\sigma}_{z}(t) \rangle_{\mathcal{N}},
\end{equation}
where $\langle \hat{\sigma_{z}}\rangle_{\mathcal{N}}$ is that in  Eq.(\ref{eq:IP}) and $\mathcal{N}=j+\frac{k}{2}$.
This series does not converge to a closed expression but it is possible to approximate it for each particular case using known methods \cite{Meystre_1974p85}.
In general, the evolution of the population inversion for an initial coherent state, Eq.(\ref{eq:EvCoh}), involves the sum of single but fixed Rabi frequencies terms,  Eq.(\ref{eq:0028}). 

\section{Particular Cases} \label{sec:Sec5}

While we believe our contribution is the identification of the underlying graded Lie algebra and the diagonalization of our model Hamiltonian, we want to show how simple is to use our results to analyze the dynamics of particular cases of our model; for more models included in our generalized JC model see Ref. \cite{Bonatsos_1993p3448}.
In all cases, Fig. X(a)	shows the time evolution of the population inversion for an initial state involving a coherent state. 
Figure X(b) and Fig. X(c) show the evolution of the boson quadratures,
	\begin{eqnarray}
	\hat{x} = \frac{1}{2} \left( \hat{a}^{\dagger} + \hat{a} \right) ~~\mathrm{and}~~\hat{y} = \frac{i}{2} \left( \hat{a}^{\dagger} - \hat{a} \right),
\end{eqnarray}
in polar plot form where the real mean value of the quadratures is the radial coordinate and time is the polar coordinate.

{\parindent 0pt \subsection{JC model}
 	Figure \ref{Fig:1} shows one of the most theoretically studied and experimentally tested models in quantum optics \cite{Jaynes_1963p89,Haroche_1985p347,Blatt_2008p1008,Wallraff_2004p162}. 
 	The JC model,
	\begin{equation}	\hat{H}_{a}=\omega\hat{n}+\frac{\omega_0}{2}\sigma_{z}+g(\hat{\sigma}_{+}\hat{a}+\hat{\sigma}_{-}\hat{a}^{\dagger}),
	\end{equation}
	allows the identification $G(\hat{n})=0$, $F(\hat{n})=0$, $f(\hat{n})=1$ and $k=1$.
	For an initial coherent state, its population inversion shows so-called collapse and revival, Fig. \ref{Fig:1}(a).
	Its quadratures show how the boson state is squeezed as time evolves, Fig. \ref{Fig:1}(b) and Fig. \ref{Fig:1}(c). 
	\begin{figure}[htb!]
		\includegraphics{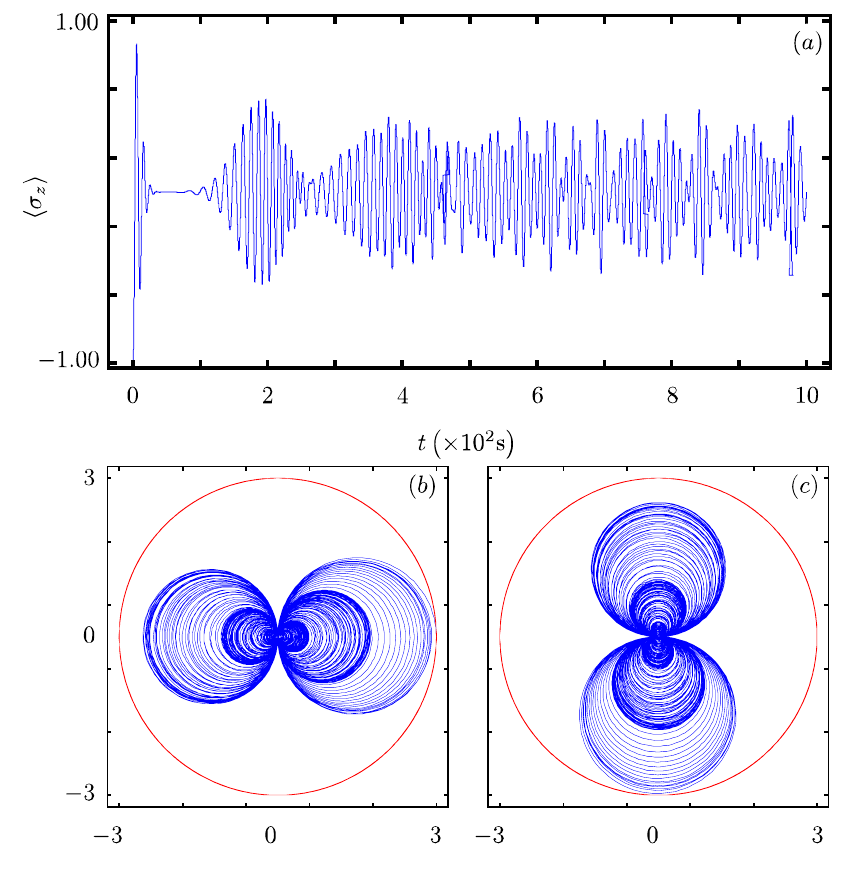}
		\caption{Time evolution of the (a) population inversion, (b) $x$-quadrature and (c) $y$-quadrature for the JC model, that is, our model with parameters $\omega= \omega_{0}$, $G(\hat{n})=0$, $F(\hat{n})=0$, $f(\hat{n})=1$, $g=0.1 \omega_{0}$ and $k=1$, for an initial state with the qubit in the ground state and the boson in a coherent state with $\alpha=3$. .}
		\label{Fig:1}
	\end{figure}	}
	
	{\parindent 0pt \subsection{JC model with intensity-dependent multi-boson coupling}
		 One of the first extensions of the standard JC model included multiboson exchange and intensity dependent coupling $f(\hat{n})=\hat{n}^{1/2}$ \cite{Sukumar_1981p211,Buck_1981p132},
	 	\begin{equation}
		 	\hat{H}_{b}=\omega\hat{n}+\frac{\omega_0}{2}\hat{\sigma}_z	+g(\hat{\sigma}_+\hat{a}^m\sqrt{\hat{n}}+\hat{\sigma}_-\sqrt{\hat{n}}\hat{a}^{\dagger m}),
		 \end{equation}
		 leading to $G(\hat{n})=0$, $F(\hat{n})=0$, and $f(\hat{n})=\sqrt{\hat{n}}$.	
		 The evolution of its populations inversion is well known for initial Fock states \cite{Singh_1982p3206}.
		 For an initial coherent state, its population inversion oscillates with a fast frequency around a value of zero, Fig. \ref{Fig:2}(a).
		 Its quadratures show that the boson state is squeezed in a faster manner and explores a more localized portion of optical phase space than in the JC case, Fig. \ref{Fig:1}(b) and Fig. \ref{Fig:1}(c).
		\begin{figure}[htb!]
		\includegraphics{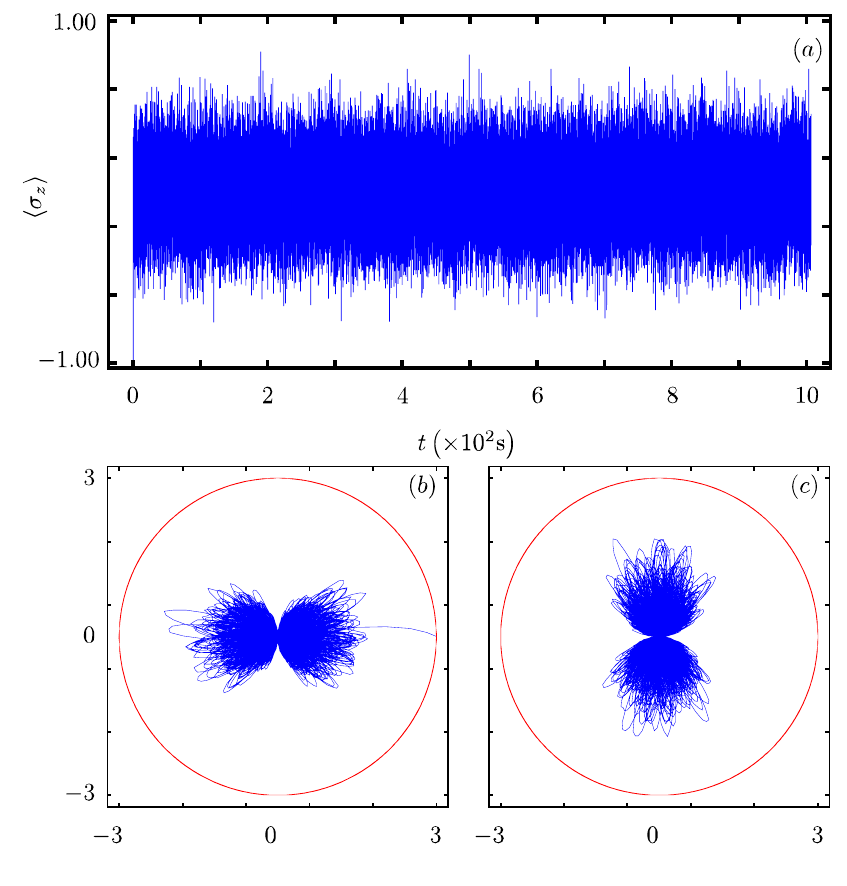}
		\caption{Same as Fig. 1 with $f(\hat{n})=\sqrt{\hat{n}}$ and $k=2$.}
		\label{Fig:2}
	\end{figure}}

	{\parindent 0pt \subsection{JC model with two-photon interaction and Stark shift} 
		This model essentially implement an additional term describing how the field intensity effects the qubit energy gap \cite{Alsing_1987p177,Nasreen_1993p1292,Gantsog_1996p445},
		\begin{equation}
		\hat{H}_{d}=\omega\hat{n}+\hat{n}\frac{\beta_2+\beta_{1}}{2}+\frac{\omega_0}{2}\hat{\sigma}_{z}+\hat{n}\frac{\beta_2-\beta_{1}}{2}\hat{\sigma}_{z}
		+g(\hat{\sigma}_+\hat{a}^2+\hat{\sigma}_-\hat{a}^{\dagger 2}).
		\end{equation}
		The parameters $\alpha$ and $\beta$ control the new features and we identify $G(\hat{n})=\hat{n}\frac{\beta_2+\beta_{1}}{2}$, $F(\hat{n})=\hat{n}\frac{\beta_{2}-\beta_{1}}{2}$, $f(\hat{n})=1$ and $k=2$.
		Its population inversion oscillates with a fast frequency and is highly localized around a value of zero, Fig. \ref{Fig:3}(a).
		Its quadratures show that the boson state squeezes in a slower manner and explores more of the optical phase space than in the JC case, Fig. \ref{Fig:3}(b) and Fig. \ref{Fig:3}(c).
	
		\begin{figure}[htb!]
		\includegraphics{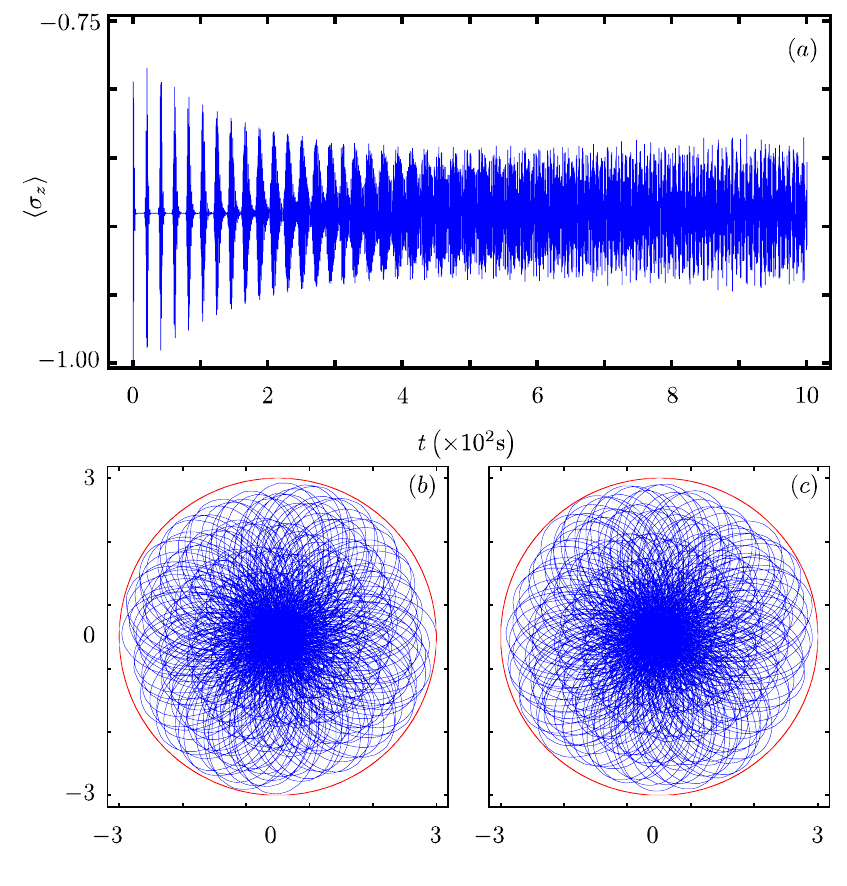}
		\caption{Same as Fig. 1 with parameters $G(\hat{n})=\hat{n}\frac{\beta_{2}+\beta_{1}}{2}$, $F(\hat{n})=\hat{n}\frac{\beta_{2}-\beta_{1}}{2}$, $k=2$, $\beta_{1}= \omega_{0}$, $\beta_{2}=0.75 \omega_{0}$.}
		\label{Fig:3}
	\end{figure}	
}
	
	{\parindent 0pt \subsection{JC model with a Kerr medium}
		 A single qubit in a single-mode cavity is surrounded by a Kerr-like medium \cite{Obada_1998p289,Gantsog_1996p445,Chumakov_1999p1817,Adanmitonde_2020p824,Cordero_2011p135502}. The medium  can be modeled as an anharmonic oscillator, the qubit undergoing
	two-photon transition is coupled to the cavity field which has a nonlinear interaction with
	the Kerr medium,
	\begin{equation}
	\hat{H}_{c}=\omega\hat{n}+\frac{\omega_0}{2}\hat{\sigma}_z+\chi\hat{n}(\hat{n}-1)
	+g(\hat{\sigma}_+\hat{a}^2+\hat{\sigma}_-\hat{a}^{\dagger 2}),
	\end{equation}
	where $\chi$ is a parameter controlling the strength of the Kerr term and we have the identification 
	$G(\hat{n})=\chi\hat{n}(\hat{n}-1)$, $F(\hat{n})=0$, $f(\hat{n})=1$ and $k=2$.\\
	Its population inversion shows that the qubit state has periodical oscillations that bring it close to the initial state for small times, Fig. \ref{Fig:4}(a). 
	Its quadratures show that the boson state also approaches its original state, Fig. \ref{Fig:4}(b) and Fig. \ref{Fig:4}(c).
	
	\begin{figure}[htb!]
		\includegraphics{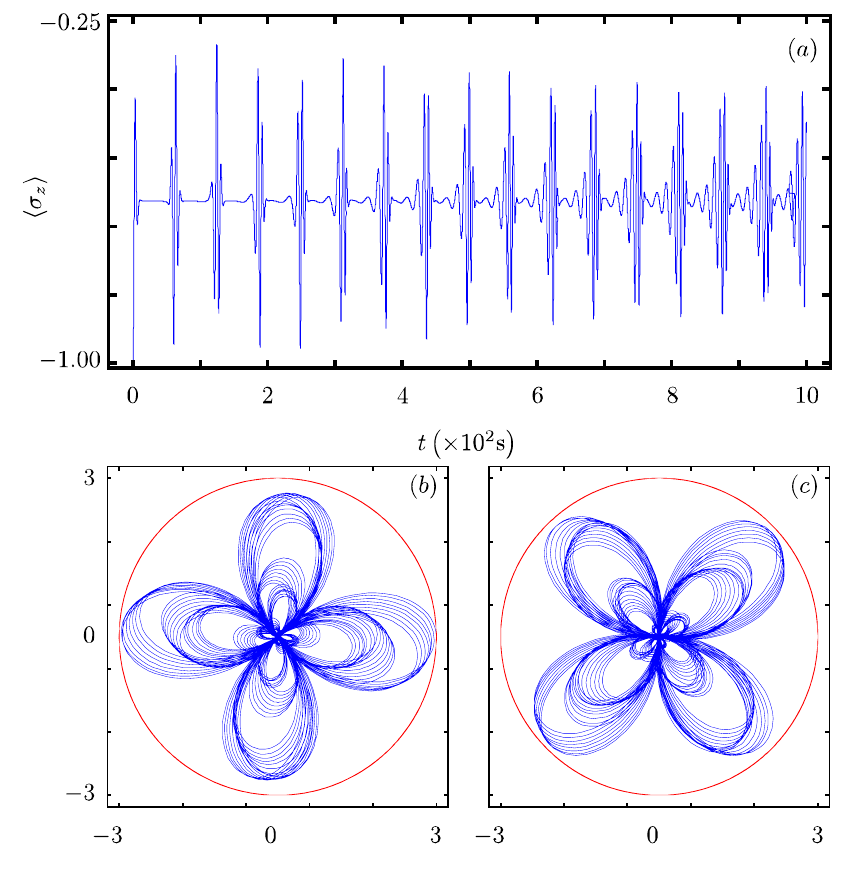}
		\caption{Same as Fig. 1 with $G(\hat{n})=\chi \hat{n}(\hat{n}-1)$, $k=2$, and $\chi=0.5 \omega_{0}$.}
		\label{Fig:4}
	\end{figure}	
	
}

{\parindent 0pt \subsection{Molecular JC Hamiltonian}
	This model arises from molecular physics or from the nonlinear Jahn-Teller effect, although long-time behavior in either case might be obscured by
	omnipresent damping  \cite{Werner_1991p4623},
	\begin{equation}
	\hat{H}_{e}=\omega\hat{n}+\frac{\omega_0}{2}\hat{\sigma}_z	+\beta\hat{n}^2+g(\hat{\sigma}_+\hat{a}+\hat{\sigma}_-\hat{a}^{\dagger }).
	\end{equation}
	The corresponding parameters are $G(\hat{n})=\beta\hat{n}^2$, $F(\hat{n})=0$, $f(\hat{n})=1$ and $k=1$.\\
	Its population inversion shows that there is almost no energy exchange between the qubit and the boson, Fig. \ref{Fig:5}(a). 
	Its quadratures show that the boson state is squeezed and explores what seems a reduced portion of optical phase space, Fig. \ref{Fig:5}(b) and Fig. \ref{Fig:5}(c).
	
	\begin{figure}[htb!]
		\includegraphics{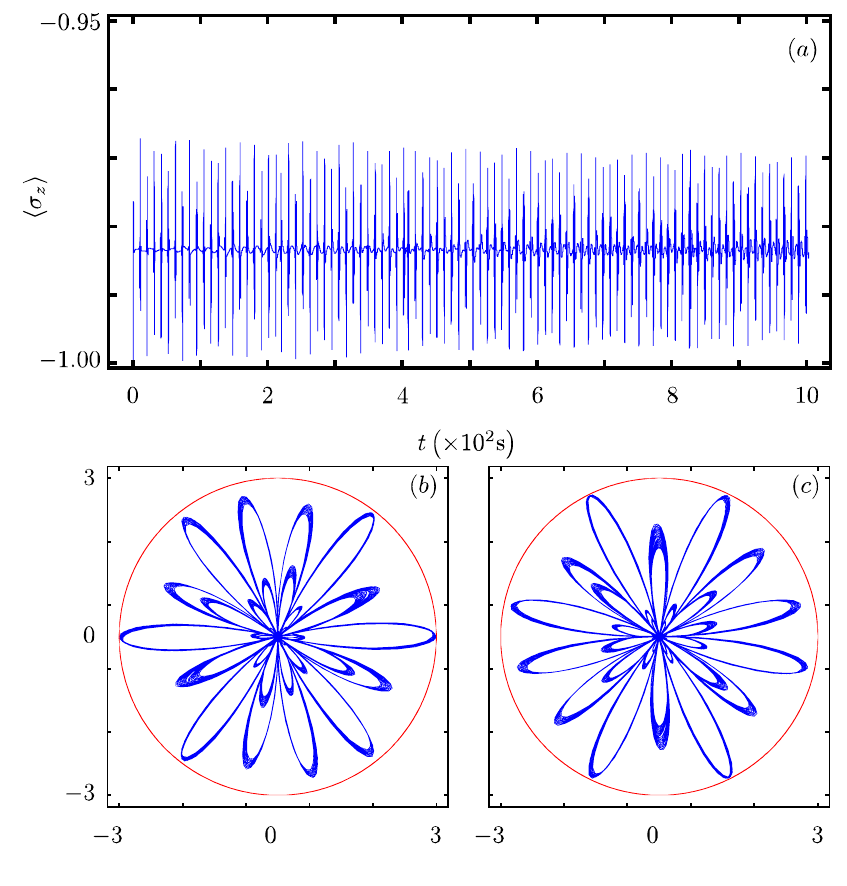}
		\caption{Same as Fig. 1 with $G(\hat{n})=\beta \hat{n}^2$ and $\beta=0.3 \omega_{0}$.}
		\label{Fig:5}
	\end{figure}
}		

  {\parindent 0pt \subsection{Algebraic JC model}
  	 Here, a deformation of the boson mode operators introduces nonlinear exchange and nonlinear boson terms  \cite{Santos_2011p015502},
	\begin{multline}
	\hat{H}_{f}=\omega\hat{n}+\frac{\omega_0}{2}\hat{\sigma}_z+\chi_a\hat{n}(\hat{n}^{\ell-1}-1)
	\\+g\left(\hat{\sigma}_+\hat{a}\sqrt{1-\frac{\chi_a}{\omega}(1-\hat{n}^{\ell-1})}+\hat{\sigma}_-\sqrt{1-\frac{\chi_a}{\omega}(1-\hat{n}^{\ell-1})}\hat{a}^{\dagger }\right),
	\end{multline}
	where the anharmonicity parameter fulfills $0\leq \chi_a \ll \omega$ and $l\geq 1$. 
	Here, we identify $G(\hat{n})=\chi_a\hat{n}(\hat{n}^{\ell-1}-1)$, $F(\hat{n})=0$, $f(\hat{n})=\sqrt{1-\frac{\chi_a}{\omega}(1-\hat{n}^{\ell-1})}$ and $k=1$.
	The population inversion presents localized oscillations around a negative value with high oscillation frequency, Fig. \ref{Fig:6}(a). 
	Its quadratures shows boson squeezing that is faster and more localized than in the standard JC model, Fig. \ref{Fig:6}(b) and Fig. \ref{Fig:6}(c).
		\begin{figure}[htb!]
		\includegraphics{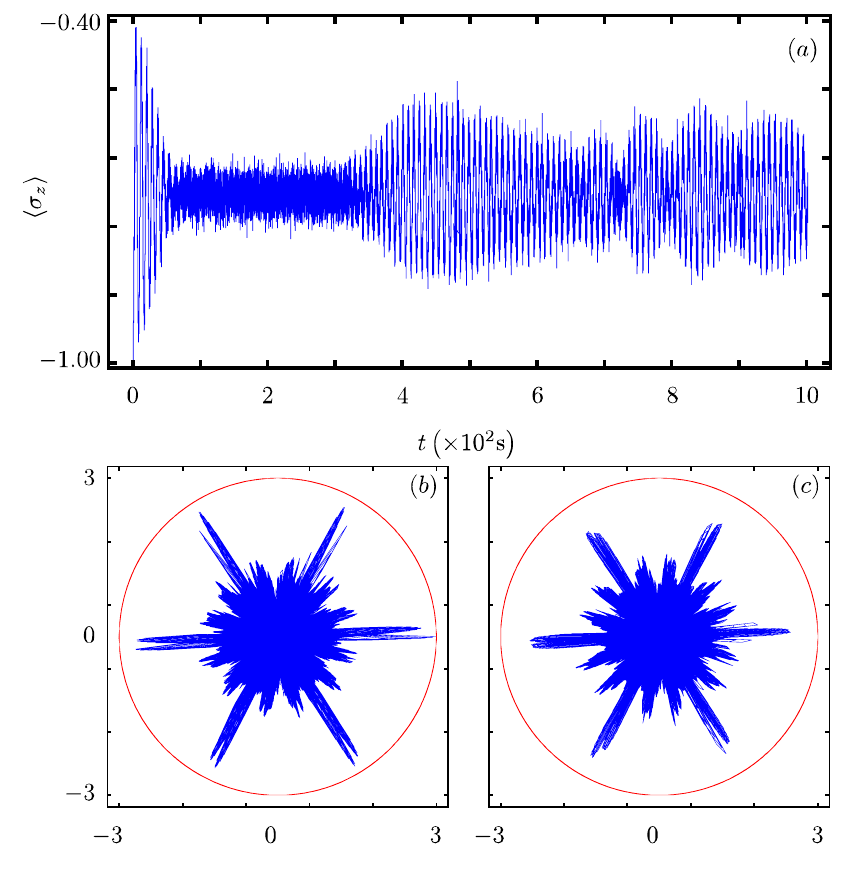}
		\caption{Same as Fig. 1 with $G(\hat{n})=\chi_a \hat{n}(\hat{n}^{\ell-1}-1)$,  $f(\hat{n})= \sqrt{1-\frac{\chi_a}{\omega}(1-\hat{n}^{\ell-1})}$, $\chi_{a}=0.5 \omega_{0}$, and $l=2$.}
		\label{Fig:6}
	\end{figure}
}
{\parindent 0pt 	
	\subsection{Parity deformed JC model}
	The parity deformed JC arises from a $\lambda$-analog of the Heisenberg algebra \cite{Dehghani_2016p38069},
	\begin{equation}
	\hat{H}_{g}=\omega\hat{n}+\frac{\omega_0}{2}\hat{\sigma}_z+\omega\lambda\left(-1\right)^{\hat{n}}
	+g(\hat{\sigma}_+\hat{a}+\hat{\sigma}_-\hat{a}^{\dagger }),
	\end{equation}
	where $\lambda$ is the deformation parameter and  $\left(-1\right)^{\hat{n}}$ is the parity operator. 
	The functions defining the model are $G(\hat{n})=\lambda\left(-1\right)^{\hat{n}}$, $F(\hat{n})=0$, $f(\hat{n})=1$ and $k=1$.
	This is a curious model as its population inversion is similar to the JC model showing a collapse and revival but localized around a negative constant bias, Fig. \ref{Fig:7}(a). 
	Its quadratures shows boson squeezing that is faster and more localized than in the standard JC model but follow a similar evolution, Fig. \ref{Fig:7}(b) and Fig. \ref{Fig:7}(c).
	\begin{figure}[htb!]
		\includegraphics{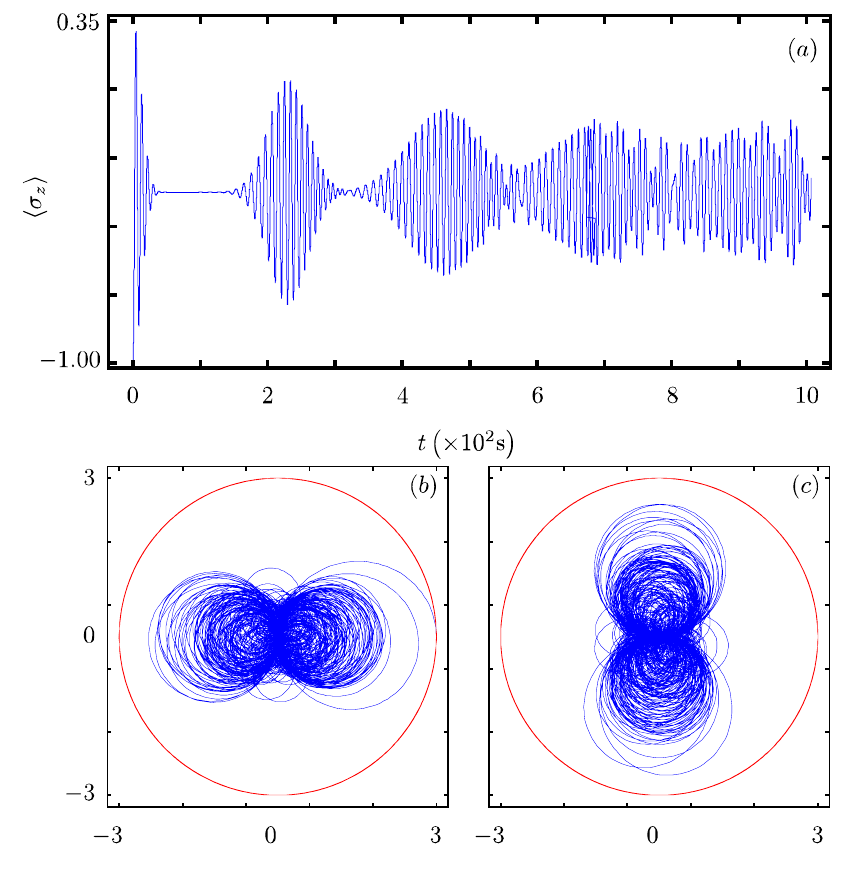}
		\caption{Same as Fig. 1 with $G(\hat{n})= \lambda \left(-1 \right)^{\hat{n}}$ and $\lambda=0.2 \omega_{0}$.}
		\label{Fig:7}
		\end{figure}
}	
{	\parindent 0pt \subsection{q-Deformed JC model}
	 This model implements deformed commutation relations for the boson operators that interpolates between Bose-Einstein and Fermi-Dirac commutation relations \cite{Chaichian_1990p980,Crnegeli_1994p1785},  
	\begin{equation}
		\hat{H}_{h}=\omega \hat{n}+\omega_{0}\hat{\sigma}_{z}+g\left(\sigma_{-}\sqrt{[\hat{n}]}a^{\dagger}+\sigma_{+}a\sqrt{[\hat{n}]}\right),
	\end{equation}
    where the deformed operator $[\hat{n}]$ is defined as
    \begin{equation}
    	[\hat{n}]=\dfrac{q^{\hat{n}}-q^{-\hat{n}}}{q-q^{-1}},
    \end{equation}
    in terms of the deformation parameter $q \le 1$.
    The corresponding parameters are $G(\hat{n}) = 0$, $F(\hat{n})=0$, $f(\hat{n})=\sqrt{[\hat{n}]}$ and $k=1$.
    Its population inversion shows a high frequency oscillation without collapse nor revival, Fig. \ref{Fig:8}(a). 
    Its quadratures shows boson squeezing that is similar to that in the standard JC model but goes faster to a reduced optical phase space region, Fig. \ref{Fig:8}(b) and Fig. \ref{Fig:8}(c).
    	\begin{figure}[htb!]
    	\includegraphics{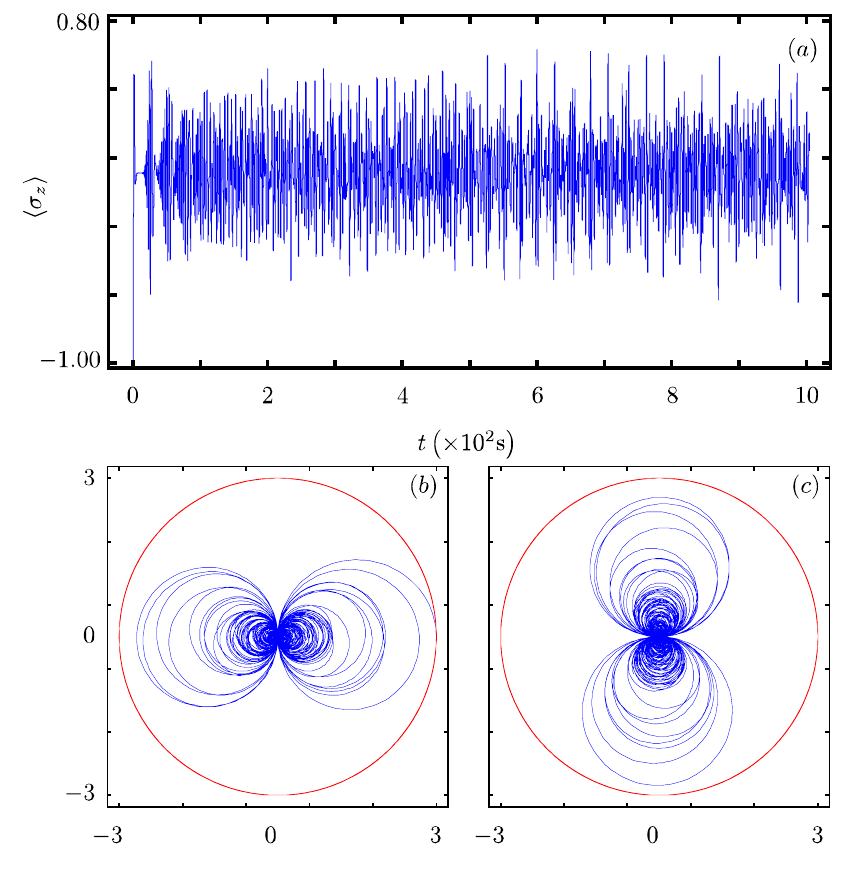}
    	\caption{Same as Fig. 1 with $f(\hat{n})=\sqrt{[\hat{n}]}$ and $q=0.9$.}
    	\label{Fig:8}
    \end{figure}
}
\section{Conclusion} \label{sec:Sec6}

We started from the well-known analogy between supersymmetric quantum mechanics and the Jaynes-Cummings model to propose an extension that includes nonlinear boson processes, nonlinear dispersive interaction, and nonlinear multiboson exchange between the qubit and the boson. 

We demonstrated that our model shows an underlying symmetry provided by a graded Lie algebra that has a similar behaviour to standard SUSY QM. 
This structure allows us to propose an operational diagonalization that provides analytic closed form eigenstates and eigenvalues, as well as time evolution.

For the sake of providing an example, we used our closed form analytic expressions to explore the dynamics of models from the literature for an initial state where the qubit is in the ground state and the boson in a coherent state. 
While a detailed analysis is not within the scope of our approach, this allowed us to identify interesting dynamics in the population inversion and in the squeezing of the boson state.
Some of these dynamics were unavailable at the time this manuscript was written.

We believe that our operational approach may open the door to the detailed theoretical and numerical analysis of extended JC models already in the literature or that may arise from current and future experimental realizations. 


\begin{acknowledgments}
F.H.M.-V. acknowledges financial support from CONACYT C\'{a}tedra Grupal \#551.
C.A.G.-G acknowledges financial support from DGAPA-UNAM postdoctoral fellowship.
\end{acknowledgments}

%
\end{document}